\documentclass[preprint,aps]{revtex4}
\usepackage{epsfig}
\usepackage{bm}
\begin{document}

\title
{\bf Compression of the current sheet and its impact into the reconnection
rate}
\author{\normalsize{S. I. Vainshtein$^1$, Z. Miki\'c$^2$, R. Sagdeev$^3$}}
\affiliation{\small \centerline{\it $^1$University
of Chicago}
\centerline{\it $^2$Science Applications International Corporation (SAIC)}
\centerline{\it $^3$University of Maryland}}


\begin{abstract}
\normalsize{
Numerical simulations of strongly compressible MHD corresponding to a stellar 
atmosphere
with substantial gravity and near force-free magnetic fields
show that the current sheet collapses (its width decreasing substantially). As
a result, the reconnection rate increases dramatically.}
\end{abstract}
\maketitle

\section{Introduction}
\label{sec0}
It is well known that magnetic configurations are dissipating
rather slowly in astrophysical conditions. Obviously, the Ohmic reconnection
time,
$$
t_O=\frac{L^2}{\eta}, $$
where $L$ is the characteristic magnetic scale, and $\eta$ is resistivity, is
huge, comparable with cosmological times. 

The dissipation becomes much more efficient in the presence of current sheets. 
In particular, the Sweet-Parker (SP) flow in the vicinity of the current sheet
results in faster reconnection rate,
so that it can be estimated as
\begin{equation}
t_{SP}=\frac{t_O}{S^{1/2}}=t_AS^{1/2},
\label{tSP}
\end{equation}
where $S$ is Lundquist number,
$$
S=\frac{C_AL}{\eta}, $$
$C_A$ is Alfv\'en velocity, 
$$
C_A=\frac{B}{\sqrt{4\pi\rho}},
$$
and $t_A$ is Alfv\'en time,
\begin{equation}
t_A=\frac{L}{C_A},
\label{tA}
\end{equation}
 see \cite{SP1}, \cite{SP2}. As $S\gg 1$ in 
astrophysical conditions, the SP
mechanism (\ref{tSP}) provides magnetic dissipation which is much faster than
Ohmic. Still, this mechanism is rather slow compared with the time of stormy
events like solar flares, or flares in the stars. 

More efficient mechanism was suggested by Petschek \cite{petschek}, with 
reconnection time,
$$
t_P=t_A\frac{8\ln{S}}{\pi}.
$$
The  Alfv\'en time (\ref{tA}) is usually small enough, so that the characteristic
time of stormy and violent events is comparable or even slightly larger than
it. For this reason, the time $t_P$ is sufficient to explain these stormy
events. It was pointed out however, that there should be some additional 
nontrivial conditions satisfied in order this mechanism to work. In particular,
there are difficulties for this mechanism  in simple two-dimensional geometry,
without additional effects like anomalous resistivity, Hall currents, localized
resistivity etc., see e.g.
\cite{biskamp1}, \cite{biskamp_book}, \cite{drake}, \cite{biskamp2}.

\section{Impact of the compressibility.}
\label{sec1}
Most of the numerical simulations and analysis are provided by incompressible
or only slightly compressible situations, although it was suggested long ago that
compressibility may speed up the reconnection rate, \cite{parker_comp}. It was 
indeed shown experimentally that the compression plays significant role in the
reconnection process, \cite{hantao}.

Previous quite  extensive numerical simulations \cite{zoran} persistently
showed that force-free compressible reconnection rate increases dramatically.
These simulations also confirm that compressibility plays key role in the reconnection
process. In particular, it was noticed that the current sheet is strongly compressed,
or one may say that it collapses, increasing the density. As a result of compression, 
the sheet width shrinks, leading to increased reconnection rate.

The compression is substantial in the low density plasmas, typical for force-free (or
near force-free) solar and stellar atmospheres, where the forces are balanced not
by a regular pressure, but rather by magnetic pressure.

The plasma and current sheet collapse  is in fact forming
a singularity, suggested by B.C. Low, \cite{low}. That was strictly one-dimensional model, and
in fact, in the vicinity of the current sheet the magnetic configuration is indeed
quasi-one-dimensional.

The simplest way to see the plasma compression is to consider an one-dimensional model,
where all the quantities depend on $x$ only, and magnetic field 
${\bf B}=
\{0,~B_y(x,t),B_z(x,t\}$. 
The force-free equilibrium can be sustained if
\begin{equation}
B_y^2+B_z^2=\rm const.
\label{equilibrium}
\end{equation}
We suggest that $B_y$ is an odd function of $x$, changing sign at $x=0$, then $B_z$
 is
an even function, and (if $B_z\ge 0$) with maximum at the origin. Suppose that the 
characteristic length of $B_y$
is $\delta$, then, according to (\ref{equilibrium}), the characteristic scale of $B_z$
would be the same. Assuming that $\delta$ is "small" (we will define this scale later), we
conclude that $B_z$ has a sharp maximum at the origin.

The equation of $B_z$ evolution is as follows,
\begin{equation}
\partial_t B_z +\nabla \cdot{\bf v}B_z=\eta\nabla^2 B_z
\label{bz}
\end{equation}
The velocity consists in fact of only one component, $v_x$, and, as seen from (\ref{bz}),
it is an odd function vanishing at the origin. Consider then (\ref{bz}) at $x=0$. 
If we look for stationary solutions, $\partial_t=0$, and, as $v_x(x=0)=0$,
we get,
\begin{equation}
B_z(x=0)\nabla\cdot {\bf v}=\eta\partial_x\partial_x B_z
\label{static}
\end{equation}
As the right-hand-side is negative (corresponding to the maximum), we conclude that
$\nabla\cdot {\bf v}<0$, i.e., compression. We now suppose that $\delta$ correspond to the
width of the SP current sheet,
\begin{equation}
\delta=\delta_{SP}=\frac{L}{S^{1/2}}.
\label{width}
\end{equation}
Then, for not very strong $B_z$, corresponding to the compressible media, we have,
\begin{equation}
\nabla\cdot {\bf v}\approx -\frac{\eta}{\delta^2}=-\frac{1}{t_A}.
\label{compress}
\end{equation}
It follows from (\ref{compress}), that, in case of SP current sheet,
 the compression proceeds with the fastest time in the
problem. 

The meaning of this compression is as follows. As mentioned above, the equilibrium
(\ref{equilibrium}) can be sustained if magnetic pressure $B_z^2/(8\pi)$ has a sharp
maximum at the origin compensating magnetic ``wall" consisting of sharp increasing of
$B_y^2$. Without this maximum of $B_z$ the configuration would collapse to the $x=0$
point (due to the $B_y^2/(8\pi)$ pressure). However, finite diffusivity smoothies out 
this maximum, not allowing the
equilibrium (\ref{equilibrium}) to survive. This plasma collapse tries to involve
magnetic field due to the frozen-in conditions, thus resulting in magnetic collapse
as well. It is this situation that was observed in previous simulations
 and also will be reported
below.

\section{The role of the gravity}
\label{sec2}
Theoretically, we can consider force-free fields (of stellar coronas), completely
neglecting the pressure, as in \cite{low}. In that case, the compressible part of
the Lorenz forces are balanced by magnetic pressure. In fact, any compression in
plasma should increase the magnetic pressure gradient to balance and eventually to
stop the compression. However, previous numerical simulations persistently showed
that magnetic pressure is not sufficient, and that the regular pressure grows.
As follows from Sec. \ref{sec1}, the magnetic pressure $B_z^2/(8\pi)$ is unable to 
balance all the forces because of the finite diffusivity, quite substantial in
SP current sheets. 

In spite of the growing regular pressure, we did not seem to find anything dramatic in
its impact. The reconnection is still much faster than in SP current sheets. In particular, it
follows from (\ref{tSP}) that the reconnection rate for SP flows is
\begin{equation}
r_{SP}=\frac{1}{t_{SP}} \sim \eta^{1/2},
\label{rateSP}
\end{equation}
and, as $\eta$ is low for astrophysical conditions (in dimensionless units), the reconnection
is inefficient. In previous simulations, with pressure included, the reconnection rate
$r \sim \eta^{0.2\div 0.3}$, i.e., much faster than the SP rate.

Still, the pressure seems to somewhat slow down the reconnection. We note, that dramatic
increase of density perturbation, and therefore increase of the pressure, is unlikely in the
stellar atmospheres. Indeed, unless there are  special conditions (satisfied, e.g., for the
solar prominences), in the presence of gravity forces, the density ``blobs" cannot be sustained
in coronas for a long time, and they are supposed to slip down along the magnetic field lines.

We expect that, if the fall down time is much less than the Alfv\'en time, then, roughly
speaking, the matter remains in stratified and almost unperturbed state all the time during the
reconnection process.

We introduce a dimensionless number,
\begin{equation}
Z=\frac{gL}{C_A^2\beta^{1/2}}= \frac{gL}{C_As},
\label{fausto}
\end{equation}
where $g$ gravity acceleration,
$$
\beta=\frac{p}{B^2/(8\pi)}\sim\frac{s^2}{C_A^2},
$$
and $s$ is the sound speed. Note that we consider low pressure case, i.e., $\beta\ll 1$.

In equilibrium state, the matter is stratified along the field lines.
We suppose that the pressure is constant on the photosphere, and therefore, in  gravitational
equilibrium, this stratification is independent of field line position, i.e., the density and
pressure are functions only of vertical dimension. 
In other words, in equilibrium state, the forces
\begin{equation}
{\bf g}\rho-\nabla p+\frac{{\bf\nabla\times\bf B}\times {\bf B}}{4\pi}
\label{forces}
\end{equation}
are balanced separately, i.e., there is hydrostatic equilibrium of a stratified
media,
\begin{equation}
{\bf g}\rho-\nabla p =0,
\label{strat}
\end{equation}
and magnetic equilibrium, i.e., force-free field,
$$
{\bf\nabla\times\bf B}\times {\bf B}=0.$$
These two equilibrium's have two different time-scales to establish. We will look for the case when
the first equilibrium has a shorter time-scale, and therefore established, while the magnetic
field is evolving slower.  

The estimated scale-height of the stratified media (\ref{strat}) is
\begin{equation}
H=\frac{s^2}{g},
\label{height}
\end{equation}
We will suppose that
\begin{equation}
Z> \beta^{1/2},
\label{weak}
\end{equation}
a condition, easily satisfied in the stellar atmospheres (like the sun). Then, 
\begin{equation}
H<L.
\label{short}
\end{equation}
Then, the recovery time for any density perturbation to vanish and to return to the
stratification (\ref{strat}) would be just the falling time on one scale-height (\ref{height}),
\begin{equation}
 t_f=\sqrt{\frac{2H}{g}}=\frac{\sqrt{2}s}{g}.
 \label{fall}
 \end{equation}

Obviously, we are interested in the case of fast established stratification, that is,
\begin{equation}
t_f\ll t_A,
\label{time}
\end{equation}
which is equivalent to the requirement
\begin{equation}
Z\gg 1,
\label{strong}
\end{equation} 
an inequality stronger than (\ref{weak}).

In this paper we are dealing only with the limiting case (\ref{strong}), leaving more general
case for the future studies. Presenting density and pressure as
$$
\rho=\rho_0+\delta\rho, $$
$$
p=p_0+\delta p.$$
Thus, in the zeroth approximation, the first two terms in (\ref{forces}) cancel each other,
$$
{\bf g}\rho_0-\nabla p_0 =0,
$$
cf. (\ref{strat}). Assuming that (\ref{time}), or (\ref{strong}) is satisfied, the
density perturbation vanishes for a short time $t_f$, being generated for a much
longer time $t_A$, according to (\ref{compress}). Therefore the density 
perturbation is weak,
\begin{equation}
\frac{\delta\rho}{\rho_0}=\frac{t_f}{t_A}=\frac{1}{Z}\ll 1,
\label{density}
\end{equation}
In the first approximation, the gravitational force is no longer balanced by the
gradient of the pressure, although they are of the same order of magnitude. Both
terms can be estimated as
$$
g\delta\rho\approx g\frac{\rho_0}{Z},
$$
while the estimation of the Lorentz force is
$$
\rho_0 \frac{C_A^2}{L}.
$$
Then, the ratio of the first two terms in (\ref{forces}) to the 
Lorentz force is
$$
\frac{gL}{ZC_A^2}=\frac{s}{C_A}=\beta^{1/2}\ll 1.
$$
Summarizing: In the zeroth approximation, the first two terms in (\ref{forces})
are canceling each other, and in the first approximation they are small compared
to the Lorentz force, if $Z\gg 1$. We will consider the
momentum equation in this approximation. As the buildup of the density and therefore
of the pressure, according to (\ref{density}), is weak, we do not expect the pressure
to impede the reconnection process, as  opposed to our previous simulations. 

\section{Formulation of the problem}
\label{sec3}
\begin{figure}
\psfig{file=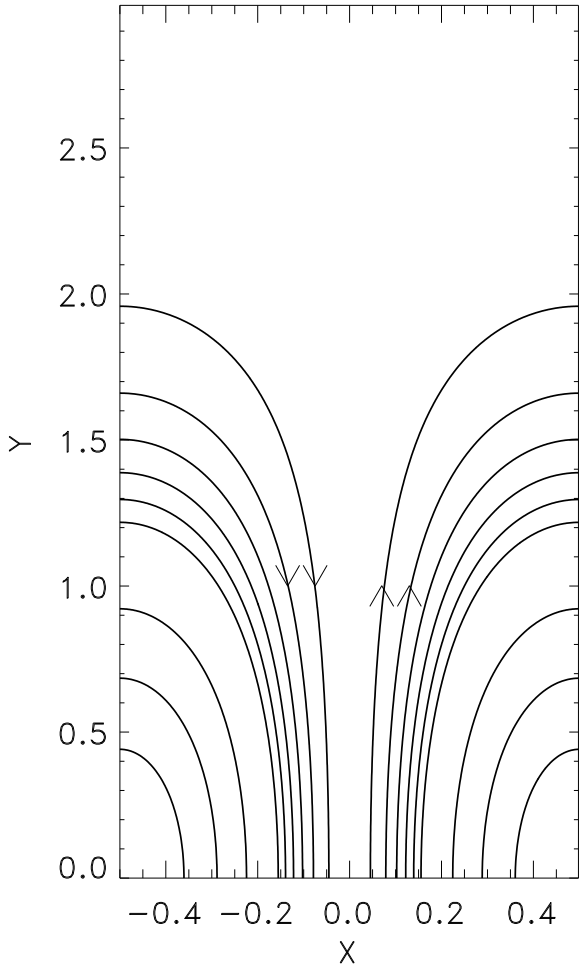,width=8in}
\caption{Initial configuration}
\label{fig1}
\end{figure}

The problem is solved in two-dimensional (compressible) MHD. That means that all
quantities depend on $x$, $y$ only (and independent of $z$), but the fields contain
all three components (the so-called 2.5-dimensional MHD).  Thus,
\begin{equation}
{\bf B}=\{{\bf B_\perp}(x,y,t),~B_z(x,y,t)\}=
\{\partial_y A_z(x,y,t),~-\partial_x A_z(x,y,t),~B_z(x,y,t)\}.
\label{magnetic}
\end{equation}
The initial
configuration is depicted in Fig. \ref{fig1}. The magnetic field is rooted on the
photosphere, $y=0$, decreasing dramatically up into the corona. The density is
stratified, i.e., it is decreasing (exponentially) from the photosphere into the
corona. 
The configuration is considered to be periodic in horizontal direction (i.e.,
in $X$-direction). We are solving the induction equation,
\begin{equation}
\partial_t {\bf B}=\nabla\times[{\bf v\times B}]+\eta\nabla^2 {\bf B},
\label{induction}
\end{equation}
with zeroth initial velocity $\bf v$. The momentum equation is presented by
Navier-Stokes equation with forcing as in (\ref{forces}), in approximation of $Z\gg
1$, see Sec. \ref{sec2}.

We can imagine that two ``sunspots" of opposite polarities are initially far away of
each other, and therefore they are not connected. Suppose that, in the course of
evolution, they approach each other, forming a configuration similar to that
depicted in Fig. \ref{fig1}. Due to frozen-in conditions, the topology of the field
lines is not easy to change, and, for a while, they remain disconnected from each
other. Due to finite conductivity, a current sheet will form, and the reconnection
starts. Further evolution is observed in our numerical simulations.

The case of $B_z=0$ corresponds to current free equilibrium. That means that, if
the pressure is neglected (which is justified in low-$\beta$ plasma (of stellar
atmospheres), the only equilibrium to which the configuration is tending to settle
down is current-free. Obviously, the configuration depicted in Fig. \ref{fig1}, does
not correspond to this equilibrium, and it has to form a current sheet, the situation
described by Syrovatsky, \cite{syr}. If $B_z\not= 0$, smooth distributed currents
(not current sheets) may be formed in corona, so that the configuration may settle
down to force-free equilibrium with (smooth) currents in the corona. This situation
is much less understood, being actually typical for stellar atmospheres, and
therefore is interesting for numerical studies.

The Alfv\'en velocity,
is small on the photosphere, and it reaches a maximum in the middle corona, and again very
small (numerically zero) is in the upper corona. For this reason, almost nothing
happens on the photospheric level, where the Alfv\'en velocity is 100 times less than
its maximum, and, for the same reason, practically everything is still well above in
corona. The reconnection is expected to take place where the Alfv\'en speed is
maximal, and it is indeed observed in the vicinity of its maximum.

Naturally, we have in mind  a real three-dimensional
configuration that can be approximated as a two-dimensional. That means that the
field in ignorable direction $z$ is eventually  also closing at the photosphere. 
In self-consistent approach, we may assume that the normal component of the velocity
vanishes on the photosphere, which means, in particular, that $v_z=0$.

The configuration depicted in Fig. \ref{fig1}, or other similar configurations with
slightly different geometry, were treated in several (at least three) different
codes, with quantitatively similar results, described in the tho next sections. 

\section{Results of numerical simulations: Measurements of the field evolution}
\label{sec4}
Previous studies were devoted to the so-called rosette-structure, in which case the
current sheet formation is inevitable, see \cite{vain}. It has been shown that the
reconnection rate strongly increases, as compared with the SP rate, \cite{zoran}, see
also the main results in Fig. \ref{fig10}. All previous simulations are dealing with no
gravity, $Z=0$, see definition in (\ref{fausto}). The new simulations are devoted to
the opposite limiting case, $Z\gg 1$, and they are performed with pseudo-spectral
code.

\subsection{Dramatic magnetic field reconstruction}
\label{sec4.1}

The main evolution of the magnetic field is described by the behavior of the
vector-potential, defined in (\ref{magnetic}). According to (\ref{induction}), the
governing equation for it reads,
\begin{equation}
\partial_t A_z +{\bf v}\nabla A_z=\eta \nabla^2 A_z,
\label{potential}
\end{equation}
which is equivalent to the $z$-component of the Ohm's law,
\begin{equation}
{\bf E} +{\bf v}\times {\bf B}=\eta {\bf j},
\label{ohm}
\end{equation}
where
\begin{equation}
\partial_t A_z =-E_z.
\label{electric}
\end{equation}

In astrophysical conditions, dimensionless value of $\eta\ll 1$, and therefore the
right-hand-side in (\ref{ohm}), is negligible, that is, in fact, the electric field
is defined from
\begin{equation}
{\bf E} +{\bf v}\times {\bf B}=0,
\label{ohm1}
\end{equation}
which corresponds to the frozen-in conditions. This is true everywhere except at the
$x$-point where $B_x=B_y=0$. Here, the typical electric field is weak, and, as seen
from (\ref{ohm}), is defined only by finite (and very weak) Ohmic decay.
\begin{figure}
\psfig{file=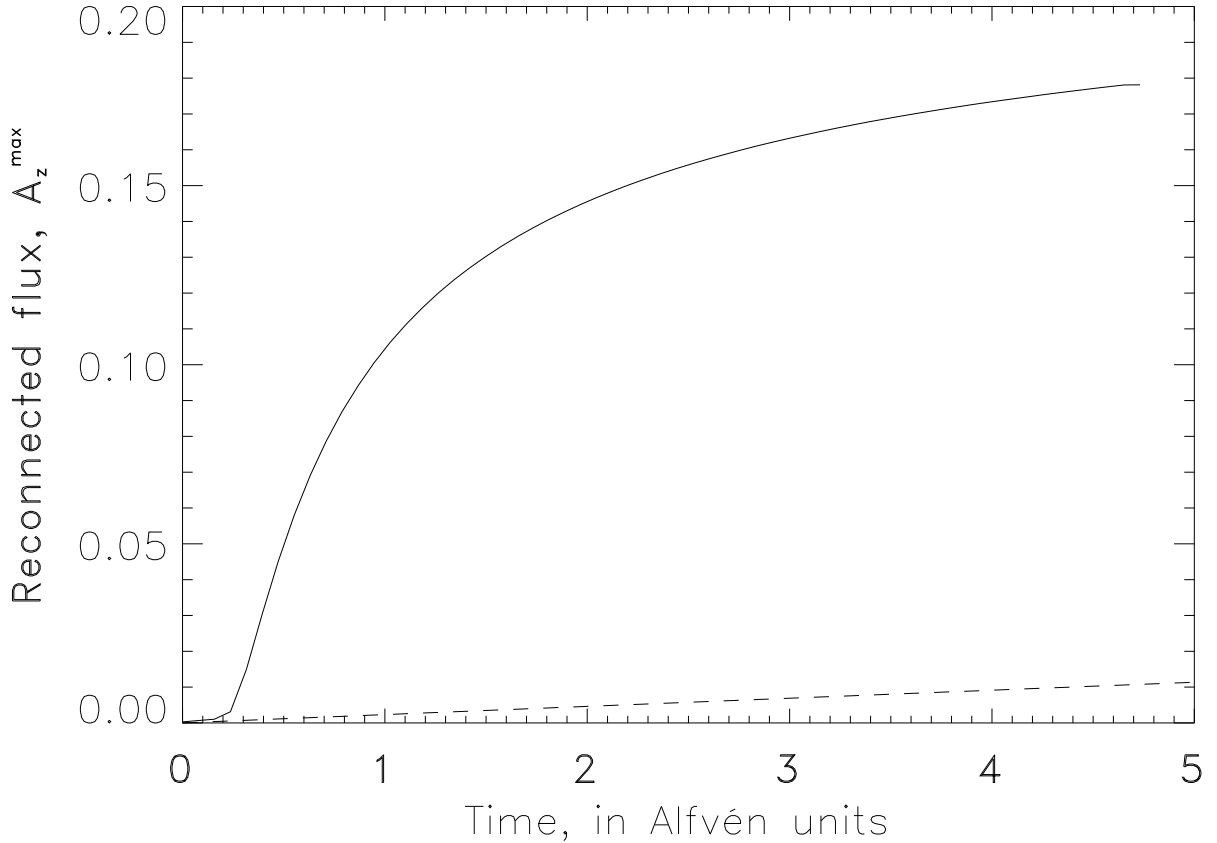}
\caption{The reconnection flux versus time; $\eta=0.00125$, for fully nonlinear case
from numerical simulations (solid line), and for pure Ohmic decay (dashed line).}
\label{fig3}
\end{figure}

The situation is dramatically different in the presence of strong currents, --
current sheets, which appear in the vicinity of the $x$-points. It is in these places
where the reconnection occurs.  In our case, the $x$-point occurs somewhere 
on the $Y$-axis, possibly moving up and down along the axis in the middle corona.

In initial configuration, Fig. \ref{fig1}, $A_z(x=0,y)\equiv 0$ on the $Y$-axis.
Afterwards the reconnection occurs, and $A_z$ grows on the axis. Almost nothing
occurs both on the photosphere, at $y=0$, and high in the corona, where the
Alfv\'en velocity is small. As a result, $A_z(x=0,y)$ acquires a maximum 
corresponding
to the $x$-point (where $B_x=B_y=0$).  The value of the maximum, $A_z^{\rm max}$,
characterizes the reconnected flux. It is important to estimate, how fast is this
maximum  growing, that is, how much flux is reconnected during a fixed amount of
time. In fact, the time derivative of $A_z$ in this point, according to 
(\ref{electric}),
defines the electric field (responsible for the particle acceleration to high
energies, etc.). In some runs, the maximum of electric field may present the
efficiency of the reconnection process. However, this happens only if this maximum is
not sharp, i.e., it is surrounded by high values of electric field - in the vicinity
of the maximum. Otherwise, the maximum is short-lived. That means that 
relatively fast reconnection did happen, but only for a short time, so that only a
small fraction of magnetic flux reconnected. For this reason, we will calculate
average electric field, which is compared with maximum possible electric field,
\begin{figure}
\psfig{file=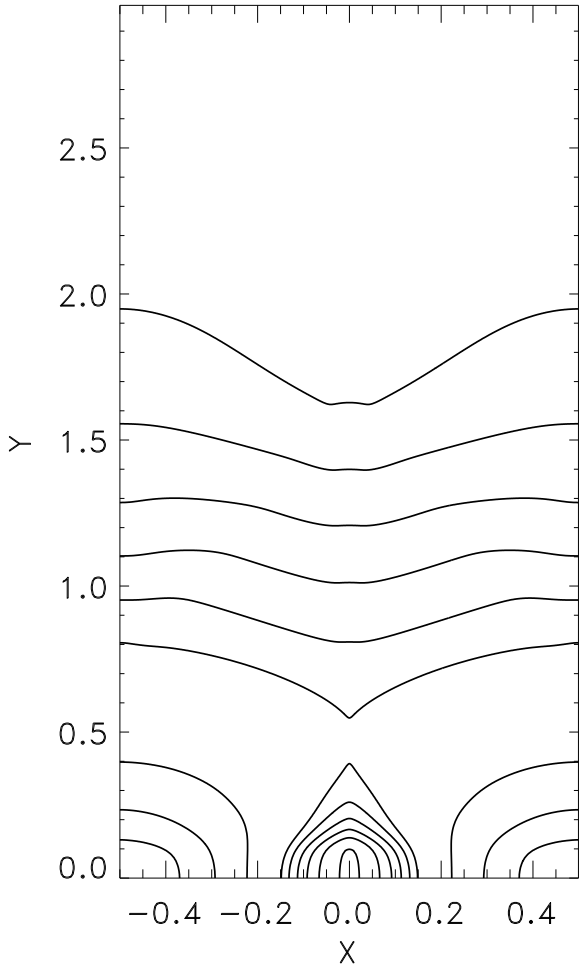,width=8in}
\caption{Final configuration, after reconnection; $t=5$ and $\eta=0.00125$}
\label{fig4}
\end{figure}

\begin{equation}
E^{\rm max}=C_A B_\perp,
\label{maximum}
\end{equation}
see (\ref{ohm1}), where the velocity was replaced by Alfv\'en velocity, maximal in
this problem.

Figure \ref{fig3} illustrates one of the runs of our simulations. It can be seen
that, for only several Alfv\'en times, a substantial amount of magnetic flux is
reconnected. This is compared with pure Ohmic decay, when all nonlinear effects are
switched off. Note that Ohmic decay (with superconductive boundary condition at the
bottom and zero at the top, and periodic conditions on the left and right), also results in
reconnection, but very slow, as seen from the figure.
\begin{figure}
\psfig{file=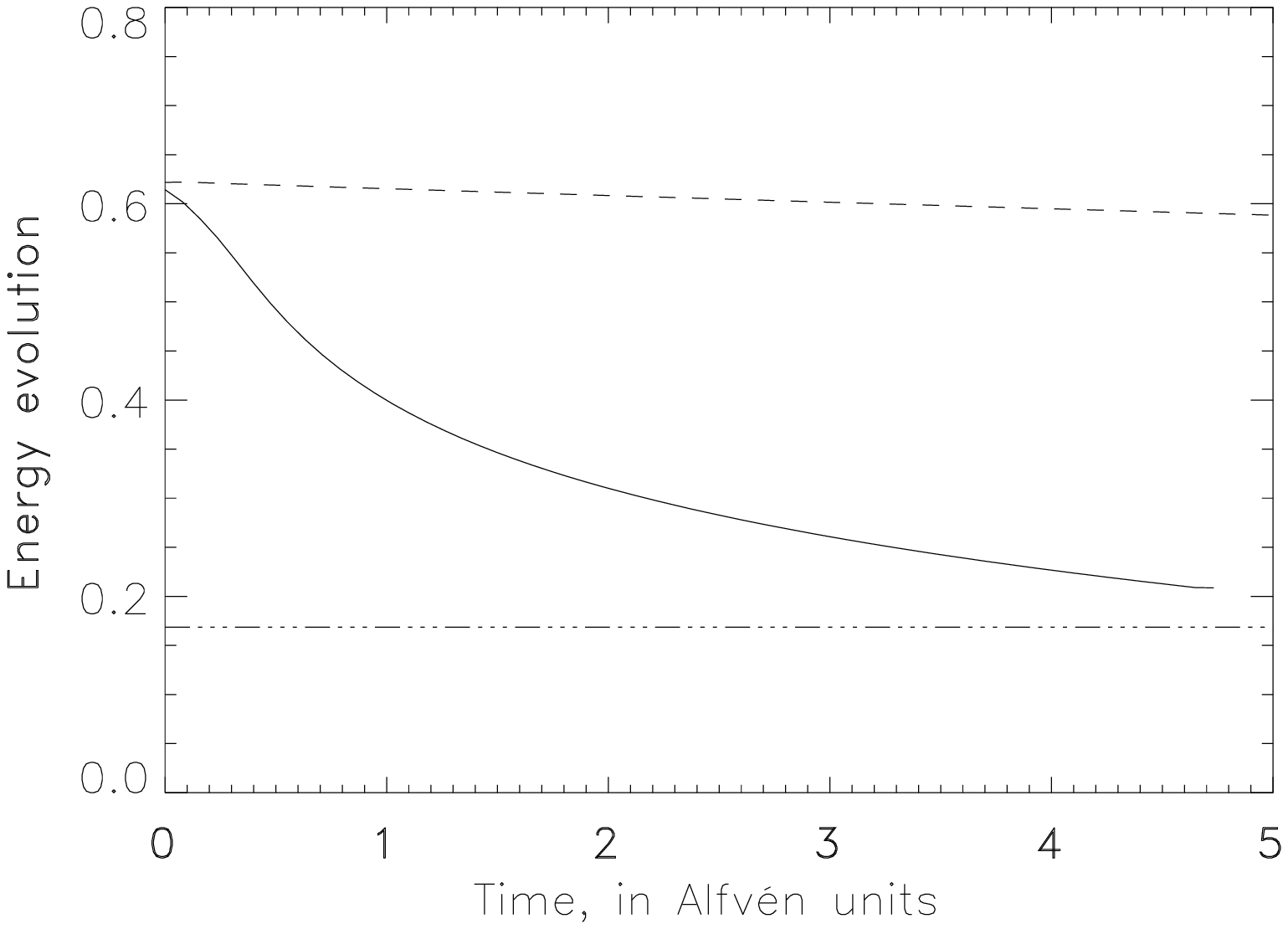}
\caption{Evolution of magnetic energy, for $\eta=0.00125$, $Z\gg 1$, depicted with a
solid line. This is compared with pure Ohmic decay (dashed line), which would
eventually end up in the lower state depicted with dashed-dotted line. }
\label{fig3a}
\end{figure}

The final state of magnetic field is depicted in Fig. \ref{fig4}, which should be
compared with initial, Fig. \ref{fig1}. It can be seen that much of the flux has been
reconnected. Note that, if all the nonlinear effects are switched off, that is, only
Ohmic decay remains, then, the configuration at the same moment of time would be
practically the same as initial depicted in Fig. \ref{fig1}. 

The magnetic energy is decreasing quite efficiently during the evolution, see Fig.
\ref{fig3a}. It is clear that the corresponding energy drop during the same time in
linear case, that is, only due to pure Ohmic decay, is much less efficient. The 
energy at $t=5$, corresponding to the configuration in Fig. \ref{fig4}, proved to be
close to the theoretical final energy, as seen from the Fig. \ref{fig3a}
(dashed-dotted line).

\begin{figure}
\psfig{file=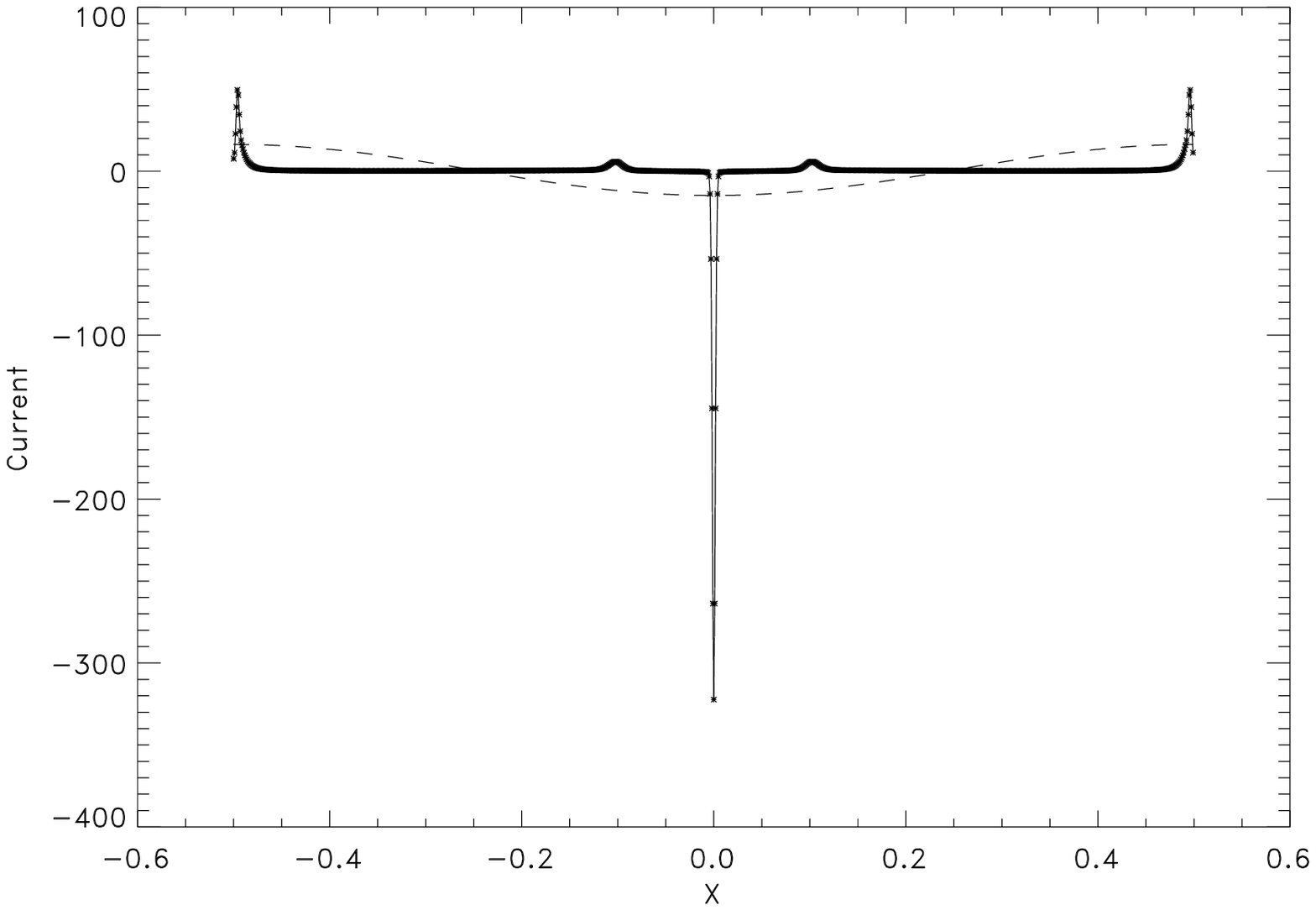}
\caption{Section of currents at $x$-point at $t=1$, $\eta=0.00125$, solid line with
asterisks. It is compared with initial current -- smooth dashed line.}
\label{fig5}
\end{figure}
\subsection{Strong currents}
\label{sec4.2}
Initial magnetic field is not at equilibrium, and, of course, the system is trying to
reach equilibrium state. This can be achieved either by current-free configuration,
for small or absent $B_z$, or force-free state, -- for finite $B_z$.
However, the initial topology of the field lines is incompatible with final
equilibrium state, see \cite{syr}, and therefore the field lines should reconnect. As
the system is driven to equilibrium by Lorentz forces, the reconnection should be
faster than Ohmic. This can be achieved if strong currents are generated, i.e., in
fact current sheets. 

As seen from Fig. \ref{fig5}, strong currents are indeed formed. An equilibrium state
would be presented by smooth currents, like the initial in the figure, but this state
is achieved only after part of the flux is reconnected, and this process of
reconnection is accompanied by strong currents.

\begin{figure}
\psfig{file=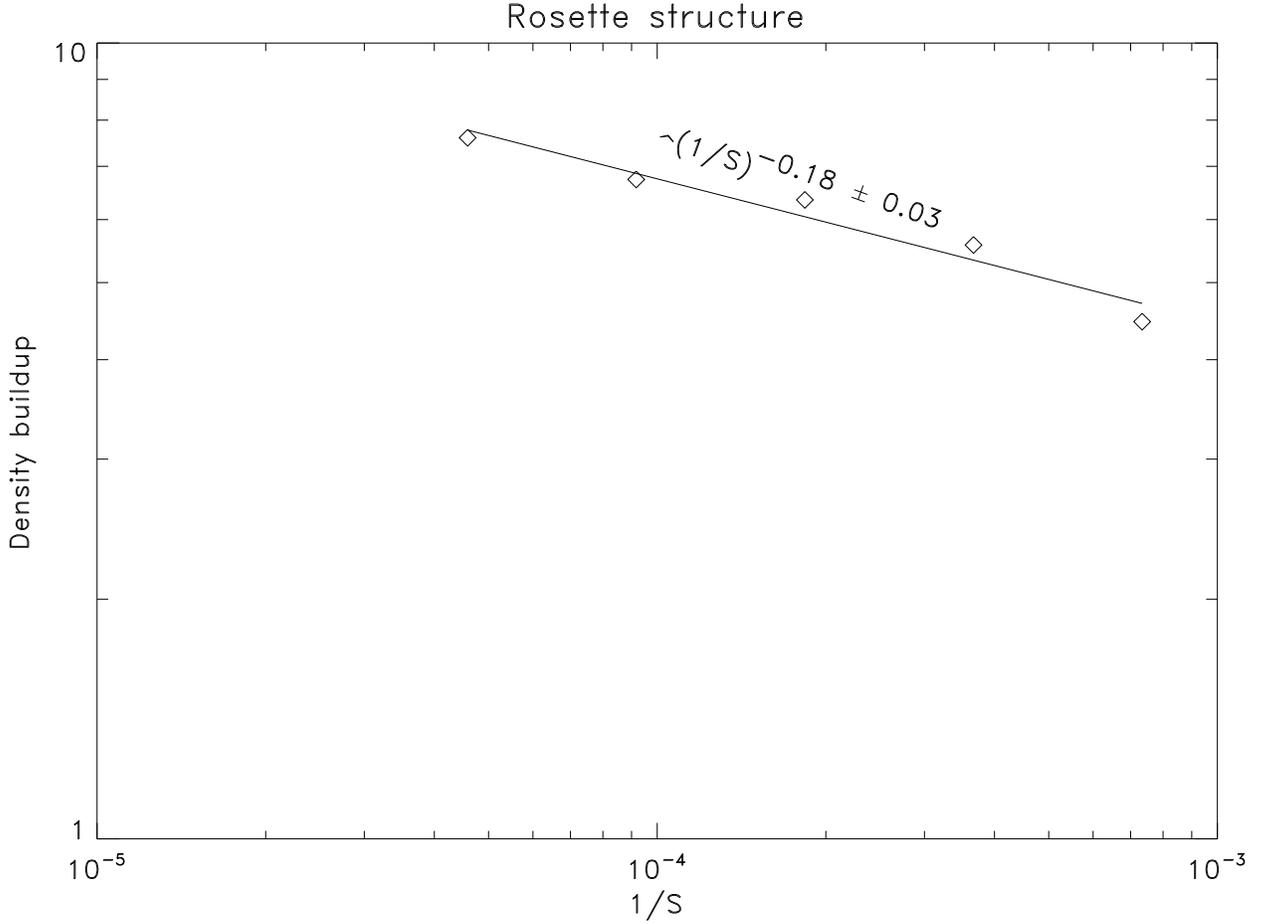}
\caption{Density buildup in case of rosette structure. Initial density is unity}
\label{fig6}
\end{figure}
\subsection{Strong compression}
\label{sec4.3}
As mentioned above, the equilibrium state can be achieved only after some flux is
reconnected. Thus, the current sheet is inevitably formed. Of course, the starting
width of the sheet is defined by SP mechanism, see (\ref{width}). In the vicinity of
the sheet the configuration is in quasi-one-dimensional equilibrium
(\ref{equilibrium}). As mentioned in Sec. \ref{sec1}, $B_z$-component acquires
maximum at the center of the sheet. For this, there should be some compression, that
is, $B_z$-component should increase approximately in
\begin{equation}
\frac{B_\perp}{B_z}
\label{init_compression}
\end{equation}
times, -- in this rough estimate we use initial characteristic values. This means
that the density would grow  the same way. For previous simulations of
rosette-structure, this estimate was $1.2\div 1.5$, or so, while the real 
compression proved to be
much stronger, see Fig. \ref{fig6}. The mechanism describing further compression
after ``natural" due to the maximum of the $B_z$-component was described in Sec.
\ref{sec1}. 
In addition,
it can be seen from Fig. \ref{fig6} that the compression grows with
decreasing resistivity with some scaling explained in Sec. \ref{sec5.1}.
 
\begin{figure}
\psfig{file=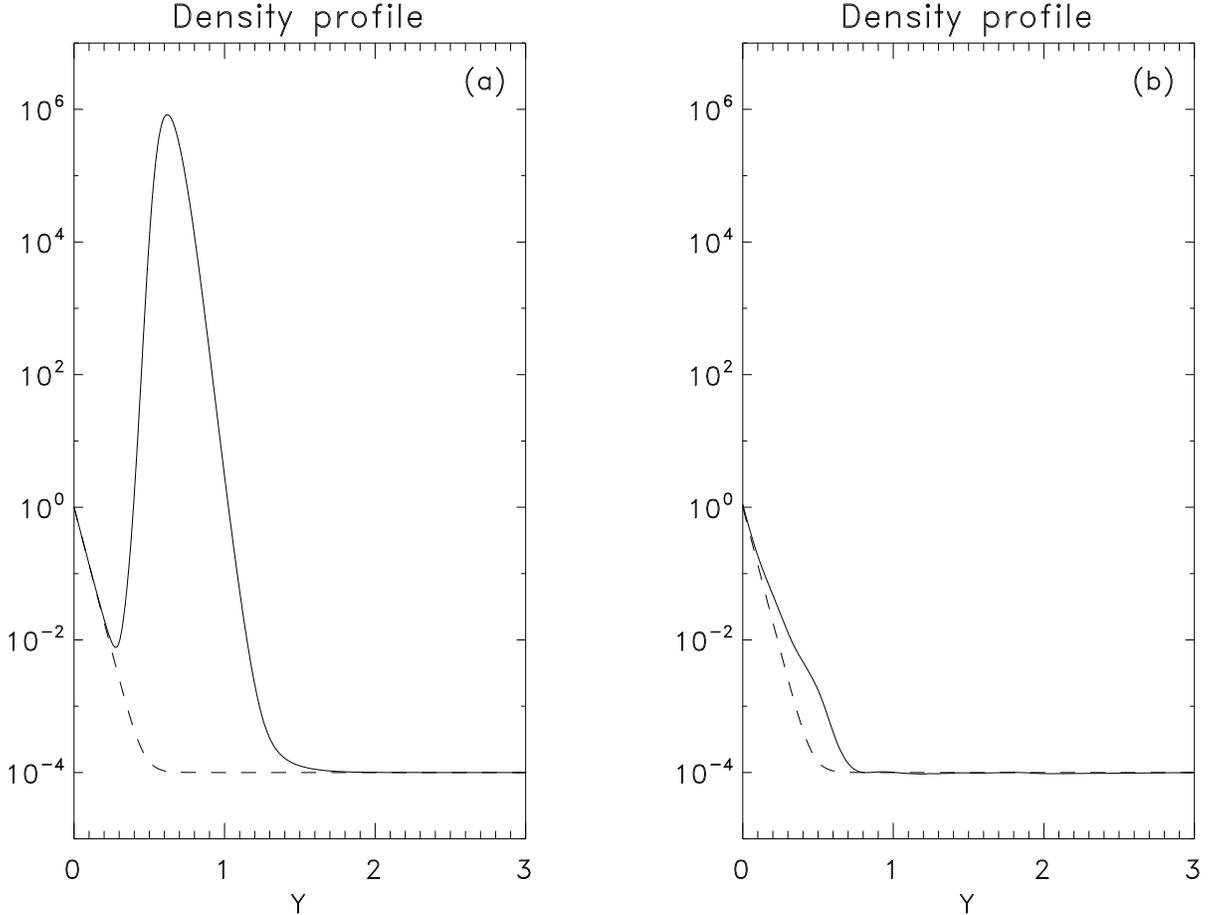,width=7.0in}
\caption{Density buildup at the $Y$-axis, depicted by solid lines. Dashed lines correspond to the
initial state. For both cases $\eta=0.00125$, and (a) is for $t=1$ and 
$B_\perp/B_z=100$, while (b) 
corresponds to $t=5$, and $B_\perp/B_z=2$}
\label{fig7}
\end{figure}

For $Z\gg 1$, the real evolution of the density is defined by the gravity forces,
and, as mentioned in Sec. \ref{sec2}, is not significant. However, we can calculate
the ``pseudo-density", that is, knowing the velocity field from the simulations, we
can calculate the density which would be real if the gravity force are switched
off. Using Lagrangian solution of the mass conservation equation, we obtain results
some examples of which are depicted in Fig. \ref{fig7}.

It is obvious that, for strongly compressible case, depicted in Fig. \ref{fig7}(a),
the compression is much stronger than ``natural", that is only $100$, while in
essentially incompressible case, depicted in Fig. \ref{fig7}(b), the compression is
not significant. We note however, that the behavior of ``pseudo-density" is only an
illustration of strong compression, as opposed to the measurements of the density
evolution in previous simulations with $Z=0$.

\section{Results of numerical simulations: The scaling}
\label{sec5}
\subsection{Ohmic decay and scaling relationships}
\label{sec5.1}
The process of reconnection is accompanied by Ohmic decay. In regular astrophysical
conditions it is quite weak, because the field is frozen-in into the plasma. The
situation is different in the presence of current sheets. The energy release during
the Ohmic decay is described as follows,
\begin{equation}
\frac{\partial}{\partial t} \int B^2 dxdy=-\eta M=
-\eta \int ({\bf \nabla\times B})^2 dxdy,
\label{energy}
\end{equation}
or, as an estimate,
\begin{equation}
\frac{1}{t_R}B^2 L^2 \approx \eta \frac{B^2}{\delta^2} \delta L,
\label{estimate}
\end{equation}
where on the right-hand-side the volume of the current sheet of the width $\delta$
is estimated as $\delta L$ because, from the numerical experiment, the length 
of
the current sheet does not change much and it is taken to be $L$.

Then, the estimated reconnection rate,
$$
\omega_R=\frac{1}{t_R}=\frac{\eta}{\delta L},
$$
or, in dimensionless form,
\begin{equation}
\Omega_R=\frac{t_A}{t_R}=\frac{L}{\delta}\frac{1}{S}.
\label{rate}
\end{equation}
For SP current sheet, $\delta$ is defined in (\ref{width}), and, according to
(\ref{rate}),
\begin{equation}
\Omega_{SP}=\frac{1}{S^{1/2}},
\label{rate_sp}
\end{equation}
cf. (\ref{tSP}).
Further compression of the sheet width, that is, decreasing of $\delta$, would result
in increase of the reconnection rate. In order to estimate this compression, we 
write (\ref{rate}) in equivalent form,
$$
\Omega_R=\frac{L}{\delta_{SP}}\frac{\delta_{SP}}{\delta}\frac{1}{S}=
\frac{\delta_{SP}}{\delta}\frac{1}{S^{1/2}},
$$
or, the compression can be defined as
\begin{equation}
\frac{\rho}{\rho_0} =\frac{\delta_{SP}}{\delta}=\Omega_RS^{1/2},
\label{compression}
\end{equation}
where $\rho_0$ is initial density. This relationship between the reconnection rate
and compression will be used below.

Note that direct measurements of the current sheet width $\delta$ could be bias. Much
more reliable are measurements of integral quantity $M$ defined in (\ref{energy}),
 the measure of Ohmic decay. According to (\ref{energy}--\ref{estimate}),
\begin{equation}
M\approx \frac{B^2L^2}{t_R\eta} \sim \Omega_R S
\label{current}
\end{equation}

\begin{figure}
\psfig{file=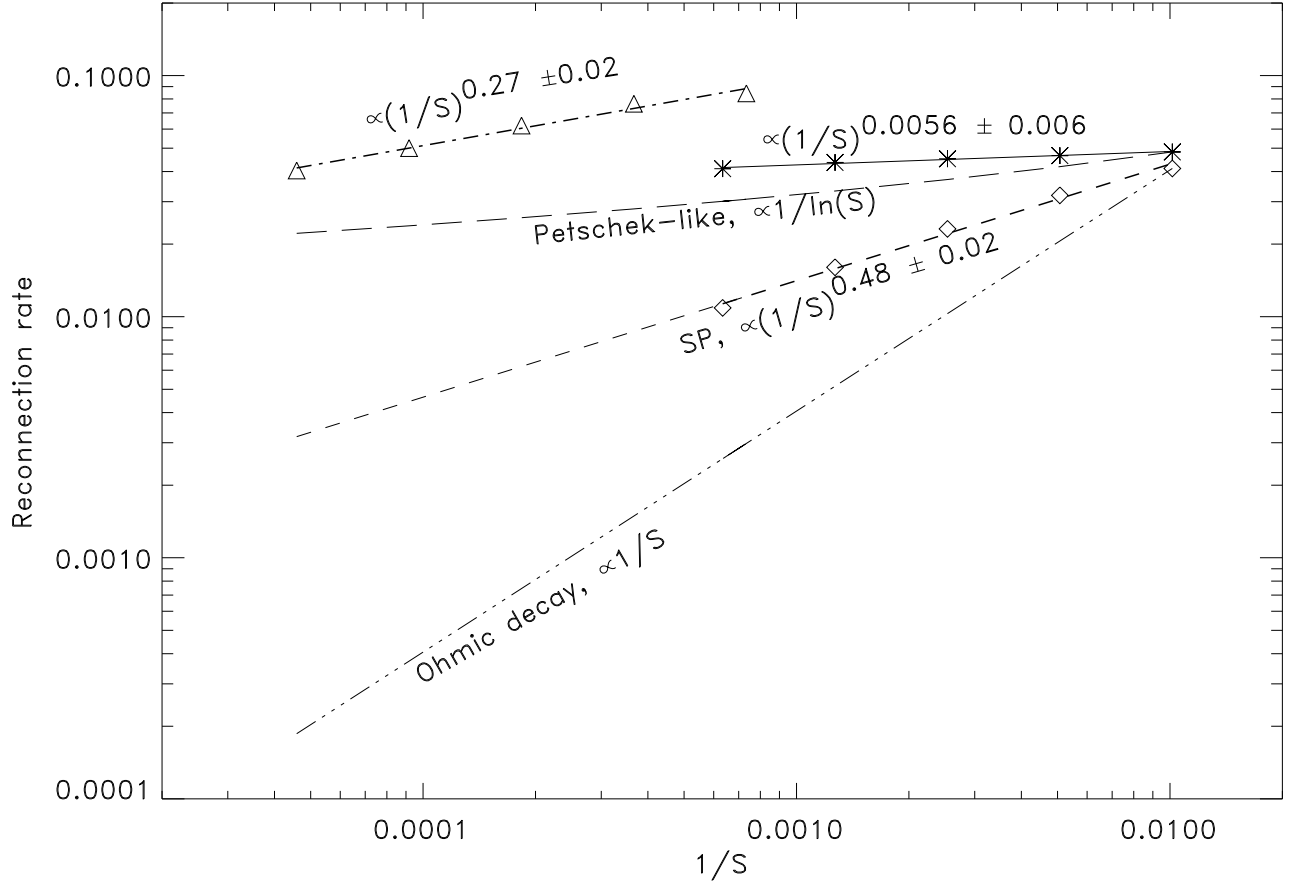,width=7in}
\caption{The reconnection rate, exponents with power-law fitting. 
Previous runs, $Z=0$, $\beta=0.01$, depicted by triangles. Asterisks correspond to $Z\gg 1$, and
$B_z/B_\perp=0.01$. The case of $B_z/B_\perp=0.4$ and $Z\gg 1$ (diamonds) is essentially
incompressible.}
\label{fig10}
\end{figure}
\subsection{Scaling laws}
\label{sec5.2}
Our main results are summarized in Fig. \ref{fig10}. Previous calculations correspond
to $Z=0$, and they definitely show reconnection faster than SP rate. All new
simulations correspond to the opposite limiting case $Z\gg 1$. If $B_z/B_\perp =0.4$
then the velocity proved to be incompressible, and then the SP mechanism is recovered
(depicted with diamonds and fitted with short dashed line). On the other hand, if
$B_z/B_\perp=0.01$, the plasma is strongly compressible, as illustrated in Fig.
\ref{fig7}. Then, the scaling (fitted with a solid line) is practically flat.
\begin{figure}
\psfig{file=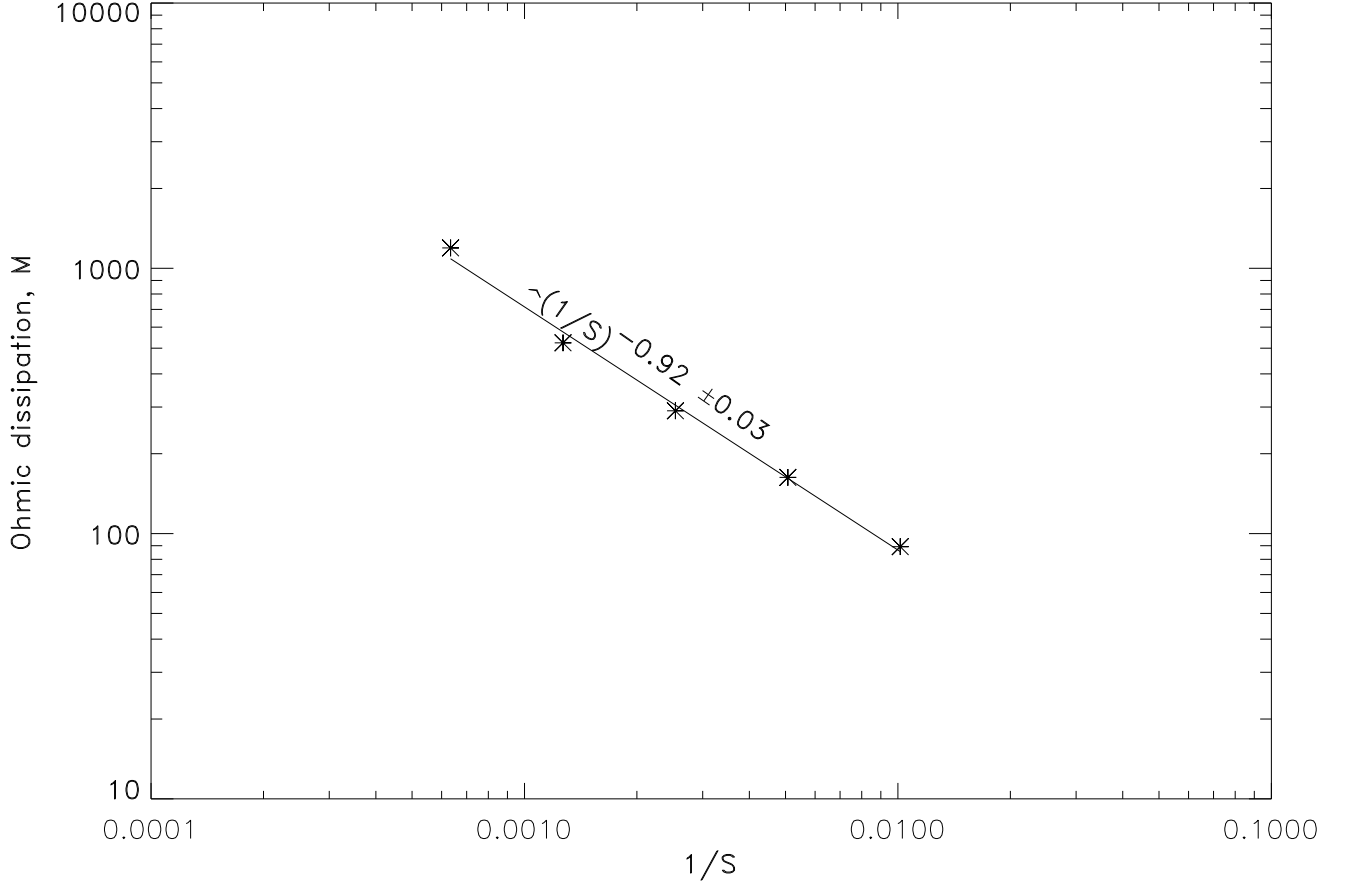}
\caption{Ohmic dissipation, $M$, defined in (\ref{energy}), as a function of 
Lundquist number $S$. Here $Z\gg 1$, and $B_z/B_\perp=0.01$.}
\label{fig11}
\end{figure}
Note that the asterisks in the figure correspond to the measurements of averaged
electric field, see Subsection \ref{sec4.1}, normalized on maximal electric field,
(\ref{maximum}). Therefore, the actual numbers should be compared with unity. It can
be seen from the Fig. \ref{fig10} that the numbers are reasonable, not very small.

Another check of the consistency of the theory is the scaling of the density 
build-up depicted in Fig. \ref{fig6}. According to (\ref{compression}), the
compression should scale as $S^{0.23\pm 0.05}$, which is consistent with the observed
scaling given in Fig. \ref{fig6}.
\begin{figure}
\psfig{file=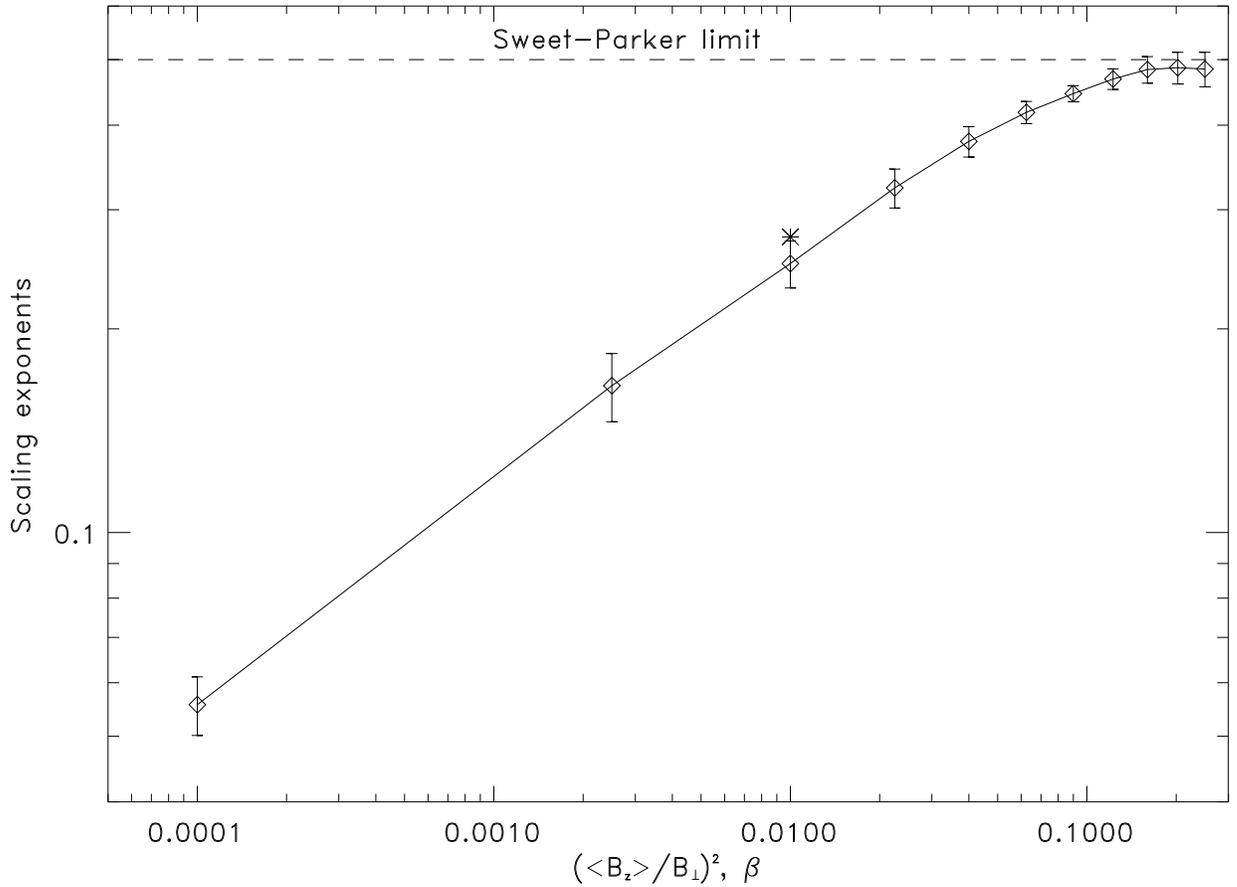}
\caption{The scaling exponents depending on compressibility. The diamonds correspond
to the different values of $(B_z/B_\perp)^2$, and $Z\gg 1$, while the asterisk
corresponds previous simulations,  $\beta=0.01$ and $Z=0$.}
\label{fig12}
\end{figure}

Finally, the scaling of Ohmic dissipation is shown in Fig. \ref{fig11}. This is 
roughly consistent with predicted in (\ref{current}), although it
does not
exactly correspond to it, presumably because all observed
scaling laws can be trusted up to the first decimal number, in spite of the
fact that the dispersion of the points is less than that, as can be seen in all 
figures.

\subsection{Scaling laws as a function of compressibility}
\label{5.3} 

We also provided numerical simulation in the same scaling range as in Fig.
\ref{fig10} for different compressible media, while all other initial parameters
(like density and magnetic field distribution, etc.) are the same as in all
simulations. The compressibility is changing with the parameter $(B_z/B_\perp)^2$.
When this parameter is small, then the media is highly compressible, while increasing
this parameter, we will approach essentially incompressible situation. As seen from
Fig. \ref{fig12}, the incompressible media is reach when $(B_z/B_\perp)^2=.04$, and
then the Sweet-Parker approximation is recovered. For higher compressibility we found
that the reconnection is faster (i.e., the scaling exponents are smaller).
Note that previously simulations correspond to moderate compressibility, in which
case the reconnection is faster than SP mechanism, but still slower than what we have
in case of high compressibility.

\acknowledgments{We thank E.N. Parker,  B.C. Low, F. Cattaneo and R. Rosner
for numerous 
discussions, and F. Cattaneo and A. Obabko for helping with simulations using
pseudo-spectral code.
This work was supported by the NSF sponsored Center for Magnetic
Self-Organization at the University of Chicago.}

\end{document}